\documentclass[nonblindrev]{informs3_hide} %

\OneAndAHalfSpacedXI %

\usepackage{etex}
\usepackage{algorithm}
\usepackage{siunitx} %
\usepackage{algpseudocode}
\usepackage{tikz}
\usepackage{amsmath}
\usepackage{graphicx}
\usepackage{multirow} %
\usepackage{booktabs}
\usepackage{array}
\usepackage{amssymb}
\usepackage{enumitem}
\usepackage{xcolor} %
\usepackage{algorithmicx}
\usepackage{float}
\usepackage{appendix}
\usepackage{longtable}
\usepackage{pdflscape}
\usepackage{subcaption}

\usepackage{natbib}
 \bibpunct[, ]{(}{)}{,}{a}{}{,}%
 \def\bibfont{\small}%

\TheoremsNumberedThrough     %
\ECRepeatTheorems

\EquationsNumberedThrough    %

\MANUSCRIPTNO{}

\begin{document}

\RUNAUTHOR{He et al.}

\RUNTITLE{Hierarchical AI Multi-Agent Investing}

\TITLE{Hierarchical AI Multi-Agent Fundamental Investing: Evidence from China’s A‑Share Market}

\ARTICLEAUTHORS{
\AUTHOR{Chujun He, Zhonghao Huang, Xiangguo Li, Ye Luo, Kewei Ma} \AFF{Faculty of Business and Economics, University of Hong Kong \EMAIL{skylar@connect.hku.hk, huangzh0624@connect.hku.hk, u3590480@connect.hku.hk, kurtluo@hku.hk, u3596913@connect.hku.hk}}
\AUTHOR{Yuxuan Xiong} \AFF{Department of Mathematics, University of Hong Kong \EMAIL{u3637747@connect.hku.hk}}
\AUTHOR{Xiaowei Zhang, Mingyang Zhao} \AFF{Department of Industrial Engineering and Decision Analytics, Hong Kong University of Science and Technology \EMAIL{xiaoweiz@ust.hk, mingyang.zhao@connect.ust.hk}}
}

\ABSTRACT{We present a multi-agent, AI-driven framework for fundamental investing that integrates macro indicators, industry-level and firm-specific information to construct optimized equity portfolios. The architecture comprises: (i) a Macro agent that dynamically screens and weights sectors based on evolving economic indicators and industry performance; (ii) four firm-level agents—Fundamental, Technical, Report, and News—that conduct in-depth analyses of individual firms to ensure both breadth and depth of coverage; (iii) a Portfolio agent that uses reinforcement learning to combine the agent outputs into a unified policy to generate the trading strategy; and (iv) a Risk Control agent that adjusts portfolio positions in response to market volatility.
We evaluate the system on the constituents by the CSI 300 Index of China's A-share market and find that it consistently outperforms standard benchmarks and a state-of-the-art multi-agent trading system on risk-adjusted returns and drawdown control.
Our core contribution is a hierarchical multi-agent design that links top-down macro screening with bottom-up fundamental analysis, offering a robust and extensible approach to factor-based portfolio construction.

}%


\maketitle

\section{Introduction}\label{Introduction}

Equity portfolio management is a multi-source, multi-horizon decision problem facing persistent uncertainty. Practitioners like fund managers synthesize macroeconomic and industry trends, firm fundamentals, price dynamics, and rapidly evolving textual information while contending with regime shifts. Signal fragility and overfitting are perennial risks when statistical models are trained on noisy, path-dependent data.
{These challenges are particularly acute for individual investors, who, as a vast body of literature documents, exhibit behavioral biases such as portfolio over-concentration and under-diversification \citep{french1991investor,huberman2001familiarity}.}
Meanwhile, risk management and governance demand transparent, attributable decisions rather than opaque end-to-end predictions. These pressures call for architectures that flexibly integrate diverse data modalities, adapt to changing conditions, and remain interpretable---ideally with a hierarchical organization that clarifies how top-level context informs downstream decisions.

Although new technologies are continually adopted in financial investing---most visibly through robo-advising and the broader deployment of machine learning, and existing work largely evaluates them in isolation. Machine learning and deep learning---such as AutoAlpha \citep{zhang2020autoalpha} and AlphaGPT \citep{wang2023alpha}---are often used for factor discovery or return forecasting; reinforcement learning (RL) is positioned as a direct portfolio policy optimizer { \citep{ye2020reinforcement}}; and natural language processing, including large language models (LLMs) is applied to sentiment extraction from news, filings, or earnings calls {\citep{xing2025designing}}. This single-technology focus yields valuable insights, but it leaves open the question of principled integration: how should macro context, structured signals, market microstructure awareness, and unstructured text be combined so that each technology operates where it adds the most value, conflicts are reconciled, and accountability is preserved? In practice, single-model pipelines centered on structured fundamentals or technical indicators often under utilize unstructured text and macro context, limiting responsiveness to news and policy shocks. Conversely, monolithic deep learning systems that ingest everything end to end can be difficult to diagnose, audit, govern in regulated environments, and often suffer from overfitting.
{ While recent LLM-based agents, such as FinMem \citep{yu2024finmem} and FinAgent \citep{zhang2024finagent} demonstrate the value of textual information, they often focus on single-stock trading or lack a comprehensive hierarchical framework for portfolio construction.}
Traditional ensemble methods typically combine signals at the feature or model level without explicit role specialization or cross-level coordination, and they can be slow to reweight under regime changes.  Our work addresses this gap with a hierarchical, role-based framework that assigns each technology to the level of the decision process where it is most effective and unifies them through adaptive aggregation and risk-aware execution.

We propose an organizing principle inspired by the structure of fundamental investment firms such as mutual funds and macro-fundamental based hedge funds: a hierarchical, role-differentiated multi-(AI)-agent system that mirrors the top-down investment process. Responsibilities of each agent are designed modular as leave-one-out ablation study is easy to carry for us to understand which component plays a bigger role in fundamental investing. The agents are aligned with real-world functions—macro strategy, security analysis, portfolio construction, and risk control—facilitating interpretability, targeted improvements, and clear attribution. Unlike the typical quant strategy that ensembles a parallel set of features by purely statistical machine learning approach, our macro-to-micro flow concentrates modeling capacity where it matters, allowing upstream agents to set context and constraints that downstream agents exploit, while preserving the ability to audit each stage independently.

At a high level, the system comprises five components arranged in a hierarchy. A \emph{Macro agent}, acting as a chief economist, occupies the top layer and analyzes macroeconomic and industry-level signals to identify sectors with favorable conditions, and therefore, allows us to focus our analyze on these particular sectors. Within these sectors, four specialized stock-scoring agents operate at the analysis layer: a \emph{Fundamental agent}, a \emph{Technical  agent}, a \emph{News  agent}, and a \emph{Report agent}, while the News and Report agents use LLMs to extract signals from unstructured text, and the other two agents depend on traditional numerical analysis. Such an approach addresses the multi-modality issue of data of different frequencies. A \emph{Portfolio agent} sits at the following up allocation layer, aggregating these heterogeneous views by learning dynamic weights over the specialized agents based on state variables such as their recent performances, thereby producing composite scores and constructing the portfolio. Finally, a \emph{Risk Control agent} forms the protective layer, adjusting exposures in response to extreme volatility. Each specialized agent can form a stand alone portfolio by investing in the top decile of its ranked list, and the hierarchical ensemble combines their strengths into a unified, adaptive allocation.

Beyond this overall design, our technical contributions center on how the hierarchy structure assembles heterogeneous technologies in a cohesive pipeline. First, we integrate LLMs for unstructured text ingestion at the analysis layer, enabling the News and Report agents to transform earnings calls, filings, and real-time news into structured signals.
{ This moves beyond using LLMs for simple sentiment analysis by embedding them in a dynamic, hierarchical workflow, addressing a key challenge in complex financial decision-making \citep{zhao2024expel,yao2023react,yao2023tree}. }
Rather than treating LLM outputs generally as features, we embed them within the hierarchical workflow to allow them to compare with the structure signals in a dynamic way.

Second, we introduce an adaptive ensembling mechanism at the allocation layer that learns dynamic weights across heterogeneous agents using previously rolling performance metrics. This online reweighting allows the system to respond to regime shifts by reallocating emphasis among fundamentals, technicals, and text-driven insights, while the hierarchical separation ensures that reweighting occurs after sector context has been established. { This RL-guided adaptive approach is similar in spirit to recent work applying RL to portfolio management \citep{ye2020reinforcement} but operates at the level of agent weights within a hierarchy.}
That is, the allocation process is guided by the instructions of the Macro agent, which resembles the real industry corporate decision process, and maintains robustness without simply collapsing signals into a single monolithic model.

Third, the hierarchy begins with a sector prefiltering stage that narrows the search space and reduces noise. By allowing the Macro agent to select favorable sectors before stock-level analysis, we dynamically focus stock pools where the signal-to-noise ratio could be higher,  and create a natural interface for incorporating macro and industry priors. This top-down selection is a key benefit of the hierarchical organization { which is grounded in financial theory, such as the well-documented industry momentum effect \citep{moskowitz1999}}: context is set once, then propagated to lower layers. Besides, compared to existing literature on applying LLM based analysis to stock level files, our approach is more scalable and less costly due to this hierarchy design.

Fourth, we enforce a clear separation of responsibilities across levels—alpha generation by specialized agents, portfolio construction via adaptive aggregation, and risk control through exposure management. This separation improves interpretability and governance by enabling attribution and drill-down diagnostics at each layer. It also supports operational robustness: components can be upgraded, swapped, or extended in a modular way without destabilizing the entire system, provided their interfaces to adjacent layers remain consistent. {This modularity satisfies key governance requirements in regulated financial environments and creates a natural interface for human-AI synergy, where human analysts may excel at interpreting intangible factors while AI handles voluminous data \citep{cao2024man}.}

Finally, we position the approach as a general framework for multi-agent coordination in financial investiment rather than a single, fixed model. The hierarchy establishes roles, interfaces, and learning rules that can accommodate new agents---such as order-book microstructure models or alternative data specialists  in the future---alongside alternative weighting schemes, including Bayesian model averaging or meta-learning, and extensions to other asset classes. In this way, the hierarchical multi-agent architecture serves as a foundation for ongoing innovation that mimics the operational structure of fundamental/value investing. In principle, this design can allow human-in-the-loop, e.g., replace our LLM-based AI Macro agent with a real experienced macro economist (i.e., a human agent), or they can operate together in a copilot form, and similar to other components of the system.

We evaluate the proposed framework using a comprehensive and challenging dataset comprising the Chinese A-share market from January 1, 2019, to December 31, 2024—a period characterized by pronounced volatility and two major market regime shifts. The dataset integrates detailed macroeconomic indicators, industry-level factors, and firm-specific features, providing a rich empirical foundation. Extensive experiments demonstrate that our framework consistently outperforms all benchmark models in both training and testing sample, while rigorous backtesting further confirms its capacity to generate robust excess returns with reduced volatility. To mitigate risks of hyperparameter overfitting and to assess generalizability, model parameters are trained exclusively on the sample from January 2019 to December 2023, with performance validated on the out-of-sample period covering January to December 2024. Additionally, an ablation study is conducted to systematically examine the contribution of each model component.

\subsection{Literature Review}

\subsubsection{Challenges in Individual Investment}
Ordinary market participants, particularly the financially unsophisticated, find it intrinsically difficult to make good investment decisions in these financial markets. Despite extensive empirical evidence in the classical finance literature \citep{badarinza2016international}. \cite{tobin1958liquidity}, \cite{markowits1952portfolio}, \cite{campbell2002strategic}, and \cite{fama2002equity} document the existence of higher expected returns (risk premiums) in equity markets. \cite{campbell2006household} finds that only a small share of people at the bottom of the wealth spectrum hold investments in publicly traded stocks.

Beyond the participation puzzle, a vast and ingenious empirical literature investigates the composition of household stock portfolios, often assuming that investors operate with partial information. Individuals tend to exhibit excessive portfolio concentration in local, domestic equities \citep{french1991investor,huberman2001familiarity} as well as their own company's stock \citep{mitchell2004lessons}. This behavioral pattern reveals significant limitations in their market comprehension and a pronounced lack of diversification knowledge.
On the contrary, our model can achieve a more comprehensive understanding of market conditions and individual stock information through hierarchical analysis at macro, industry and individual stock levels, while integrating multiple data sources including news, financial statements, equity reports, and price-volume data, thereby enabling the selection of high-quality stocks from a larger stock pool.

A further critical shortcoming of individual investment strategies is suboptimal portfolio construction, manifested as persistent under-diversification. A robust empirical finding indicates that retail investors tend to allocate a disproportionate share of their capital to a concentrated set of risky assets, typically individual equities or poorly-diversified fund products; see \cite{barber2000trading}, \cite{gargano2018does}, and \cite{d2019promises}.
Our Risk Control agent enables more effective risk mitigation in investment decision-making. In back-testing results, during periods of adverse market conditions, the model demonstrates significantly lower drawdown magnitudes compared to the benchmark.

\subsubsection{Limitations of Human Financial Advisors}
Although some may argue that investment advisors can enhance portfolio returns, there is in fact scant empirical evidence in the literature to support this claim.
\cite{gennaioli2015money} point out that professional financial guidance can help address underdiversification issues and enhance investment performance for individual investors. Providing empirical evidence that greater trust in financial advisers leads to increased risk-taking among investors. However, the higher returns experienced are insufficient to offset the elevated fees. This suggests that investors may either be unaware of the costs associated with advisory services or prioritize factors beyond maximizing portfolio returns in their interactions with financial advisers.

Contrary to the conventional expectation that human investment advisors contribute to superior investment performance, numerous studies indicate that human investment advisors confront significant practical limitations. Recent empirical evidence suggests that conflicts of interest inherent in the advisor-client relationship can lead to a significant distortion of portfolio asset allocation.
Investment accounts managed by financial advisors and mutual funds sold through brokers tend to deliver poorer performance compared to self-managed portfolios \citep{bergstresser2008assessing,christoffersen2013consumers,chalmers2020conflicted}.
\cite{hackethal2012financial} demonstrate that advised accounts under-perform self-managed portfolios in net returns and risk-adjusted performance (Sharpe ratios), particularly under bank advisors, with elevated trading costs from commission-driven turnover. \cite{linnainmaa2021misguided} find that financial advisors exhibit investment behaviors similar to their clients, characterized by frequent trading, return chasing, a preference for expensive actively managed funds, and under-diversification. Investment advisors tend to accommodate the behavioral biases of their clients, prompting them to pursue strong historical performance and invest in actively managed mutual funds \citep{mullainathan2012market}.
Retirement plan administrators tend to prioritize their own proprietary funds when creating investment lineups \citep{pool2016pays}. Client behaviors strongly correlate with advisors' personal strategies \citep{linnainmaa2021misguided}. In contrast, LLM-derived analytical results can circumvent the subjective biases inherent in human advisory services and remain uninfluenced by personal preferences or individual predispositions.

As a consequence, many individuals exhibit pronounced mistrust towards human financial advisors, driven by apprehensions about being subjected to financially motivated malfeasance \citep{calcagno2015financial,lachance2012financial,burke2021trust}.

As \cite{reher2024robo} note, human investment advisors typically serve an exclusive clientele of affluent individuals. In contrast, our system is designed to deliver cost-effective advisory services to a much broader demographic.

\subsubsection{Rise of Robo-Advisors}
The advent of robo-advisors in the mid-2000s was driven by the shortcomings of traditional financial advisory services.
\cite{d2019promises} offer empirical evidence that robo-advising tools enhance portfolio diversification, mitigate behavioral biases, and improve investment performance.
\cite{d2021robo} investigate the emergence of robo-advisors, analyzing their classifications, benefits, and challenges while outlining unresolved interdisciplinary issues that will determine the evolution of algorithmic financial advice.
Robo-advisors are capable of serving individuals with significantly lower wealth levels, while human financial advisors, being limited by time, generally focus on catering to more affluent households \citep{reher2024robo}.

In the study of an Automated Financial Management service, \cite{reher2019automated} finds that portfolios constructed through the service exhibit a higher degree of diversification compared to those that are self-managed. Furthermore, they document that a reduction in the minimum investment threshold required for access is associated with a significant increase in customer fund inflows.
Moreover, in contrast to traditional human advisors who use survey-based approaches to assess investor risk preferences, robo-advising can leverage machine learning algorithms to infer risk tolerance from investors' historical investment decisions \citep{alsabah2021robo}, especially for financially unsophisticated investors.

Several studies compare robo-advisors with human analysts. \cite{coleman2022human} compares algorithmic (``Robo'') analysts to their human counterparts, finding that robo-analysts produce less optimistic and more frequently revised recommendations with reduced conflicts of interest. Their automated processing of complex disclosures generates long-term investment value, significantly outperforming human analysts' buy recommendations. Similarly, \cite{cao2024man} explores the synergy between human analysts and AI in stock return predictions, noting that while AI excels in handling voluminous data, humans are better at interpreting intangible assets and financial distress. Combined approaches reduce extreme errors, suggesting complementary strengths.

Recent research examines the profiles of early adopters and the industry-wide effects of automated financial advisory platforms.
Based on the FINRA 2015 survey data, \cite{kim2019robo} and \cite{lu2023bbt} find that younger age groups, higher disposable income, and greater risk propensity are significantly associated with early adoption of automated financial advisory platforms. Similarly, \cite{baulkaran2023uses} examines robo-advisor users in India, revealing that typical users are young, male, married, small investors, and professionals. \cite{ben2022mutual} further notes that robo-advisors significantly reduce demand for human financial advice, especially among distrustful investors, underscoring their disruptive potential in wealth management.
Our framework exhibits strong interactivity and can articulate analytical reasoning and rationale, thereby enhancing accessibility and usability for investors, particularly elderly and non-professional investors.

\subsubsection{LLMs for Financial Decision-Making}

Substantial research is dedicated to creating versatile LLM-based agents capable of sequential decision-making \citep{zhao2024expel,yao2023tree,yao2023react}.
Moreover, scholars have begun to investigate strategies for leveraging LLM agents to achieve enhanced performance in more complex decision-making tasks within the financial domain
\citep{de2023llm,zhao2024revolutionizing,liu2025fmdllama,cao2024risklabs}, wherein environments exhibit greater volatility, resulting in a multitude of unpredictable factors that hinder the agent's capacity for precise introspection into the causes of suboptimal decision outcomes. FinMem \citep{yu2024finmem} augments performance in single-stock trading by integrating a memory module with its LLM agent to facilitate a cycle of reflection and refinement. FinAgent \citep{zhang2024finagent} enhances trading profitability by leveraging an external quantitative tool to counteract market volatility. AlphaGPT \citep{wang2023alpha} and AutoAlpha \citep{zhang2020autoalpha} to discover both price-volume and formulaic alpha factors, further underscoring their adaptability across financial applications.

Diverging from conventional rule-based and RL methodologies, which are often constrained to using only price data and typically function as ``black boxes'' lacking interpretability, our framework leverages a multi-source data integration approach. It provides not only stock recommendations but also the explicit rationales behind them, significantly enhancing the model's transparency and explainability.

Furthermore, we address the limitations observed in current mainstream models. For instance, models like TradingGPT  often focus on single-stock trading tasks and are validated through backtests on a single asset. Similarly, models such as FinAgent and FinCon  base their backtesting on a small selection of mainstream stocks. In stark contrast, our framework conducts stock selection and backtesting across a much broader and more representative stock pool—the CSI 300 index—thereby yielding results with substantially greater robustness and generalizability.

While some advanced models like FinCon have begun to incorporate portfolio management, our framework advances this concept further by constructing a hierarchical portfolio that allocates capital from the macro level down to individual stocks. Critically, it adopts a RL mechanism to dynamically adjust the weights among different agents. This process systematically amplifies the influence of agents with superior historical performance in the portfolio construction process, ensuring continuous optimization and adaptation.

Compared to the open-source MASS \citep{guo2025mass}, a multi-agent scaling simulation framework designed for portfolio construction, this model demonstrates superior performance in backtesting on the CSI 300 index.

\smallskip

The remainder of this paper is organized as follows. Section~\ref{sec:methodology} outlines a detailed description of our proposed methodology, multi-agent system architecture, and the key agents that deals with different kinds of data.
Section \ref{sec:experiment} introduces data, conducts a comprehensive comparative analysis, benchmarking our model against both standard market indices and other contemporary multi-agent models to demonstrate its efficacy and robustness. Section \ref{sec:conclusion} concludes.

\section{Methodology}\label{sec:methodology}
\begin{figure}[ht!]
\FIGURE{
    \includegraphics[width=0.9\textwidth]{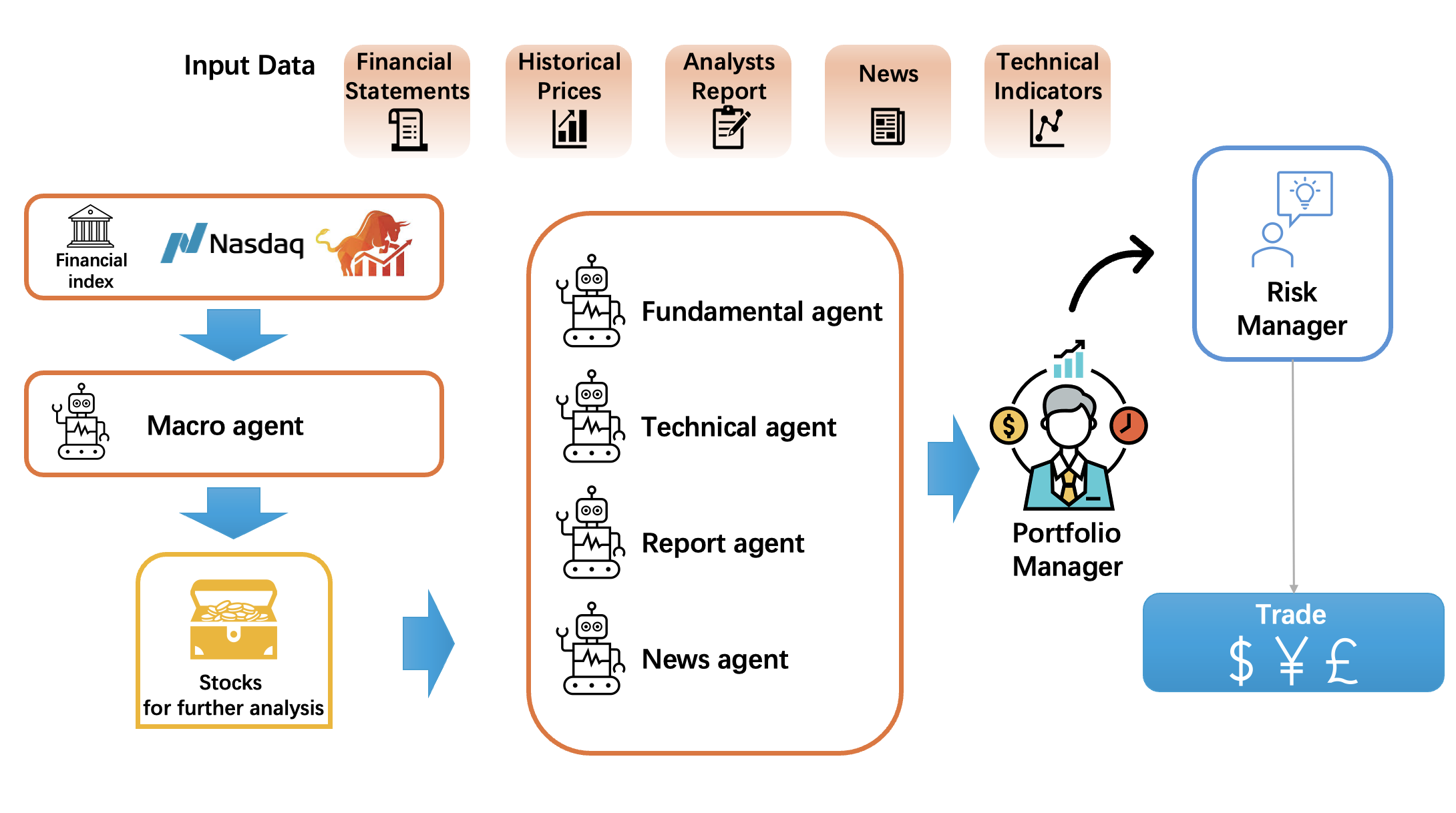}}
{Multi-Agent Trading System \label{fig:AI-trading}}
{}
\end{figure}

Our system implements a QuantMental investment framework, following a top-down framework: from macro and industry to the firms analysis; see Figure~\ref{fig:AI-trading}. At the top level, the Macro agent identifies the most promising industries, serving as a filter to focus subsequent analysis. Within the selected industries, multiple firm-level agents—including the Fundamental agent, Technical agent, News agent, and Report agent generate distinct scores for each stock, capturing diverse aspects such as financial health, market behavior and informational sentiment. The Portfolio agent then dynamically allocates weights across these firm-level signals, optimizing for expected Sharpe ratio and annual return while ensuring diversification between agents. Finally, the Risk Control agent adjusts overall portfolio exposure according to market volatility, helping the system navigate extreme events such as policy shocks or geopolitical crises. By integrating top-down sector analysis with multi-agent stock evaluation and adaptive portfolio management, the system aims to create a prevailing portfolio and manage risk in a dynamic market environment.

\subsection{Data Flow and Obfuscation}
The proposed fundamental investing workflow combines a CSI~300 stock pool with industry-level filtering mechanisms and firm-level multi-agent evaluation.
This integrated approach enables the systematic incorporation of multi-level market information, encompassing macroeconomic conditions and industry-level characteristics, firm-specific features, and benchmark-based variables derived from equity indices.

We curate the stock pool from the Chinese A-share market, the CSI 300 index, $\mathcal{S}_t^{\mathrm{CSI}} \subseteq \{1,\dots,n\}$ at rebalancing date $t$. Compared to existing literature on portfolio construction and asset selection frequently suffers from limited scope, either by confining analyses to short time intervals or by focusing exclusively on a small group of high-performing U.S. equities (e.g., Google and Apple), it provides a broader and more representative analysis.

We accord with the China Securities Regulatory Commission (CSRC) Industry Classification for the industry-level features \(
\mathbf{S} \in \mathbb{R}^{T \times d_s}
\) over $T$ trading days with \(d_s\) features. We collect industry daily return and macro data including CPI, money supply and PMI, so the Macro agent facilitates comprehensive analysis in both macro and industrial perspectives.

The firm-level scoring agents, which incorporate perspectives in parallel from fundamental, technical, news, and analyst report dimensions, operate on firm-level features \(
\mathbf{F} \in \mathbb{R}^{N \times T \times d_f}
\) over $T$ trading days with \(d_f\) features and $N$ assets. The historical price data encompass daily market observations, including opening, high, low, closing, and adjusted closing prices and trading volume. Technical indicators complement the market dynamics with Moving Average Convergence Divergence (MACD), Relative Strength Index (RSI), and Stochastic Oscillator (KDJ)—computed per asset to identify trends, momentum, and potential reversal points. We curated daily news feeds are aggregated from Google and Baidu covering the informational context about the company. Analyst reports offering professional assessments, including investor activities, legal dishonesty details and the company financial activities.

To safeguard sensitive financial information and mitigate the risk of training leakage within the backbone LLM, the firm-level features are deliberately obfuscated prior to their utilization by the scoring agents. This obfuscation process conceals key identifiers and proprietary signals embedded within the raw data, ensuring that the model does not inadvertently internalize asset-specific knowledge that could create unintended information leakage.

After collection and obfuscation, the data are transmitted directly to the agents in order to prevent distortions common in chain-based communication. Such distortions often arise when numerical values are abstracted into summaries, leading to rounding errors, omission of key figures, or even fabricated values. The proposed approach preserves data fidelity, ensuring agents receive accurate and reliable information.

\subsection{Macro Agent: Industry-Level Filtering}

From the basic stock pool, the Macro agent selects industries by integrating
macroeconomic regime signals with industry-level momentum. Building on the Merrill Lynch investment clock \cite{merrill2004}, the industry rotation framework \cite{grauer1990}, and the documented industry momentum effect \cite{moskowitz1999}, we classify the economic cycle into four phases: \emph{recovery}, \emph{overheating}, \emph{stagflation}, and \emph{recession}. Industry allocation is determined by a combination of prevailing macroeconomic conditions and systematic industry-level momentum effects, with portfolio adjustments incorporating liquidity conditions.

We use four macroeconomic indicators:
\begin{itemize}
    \item M1 Money Supply:
    Let
    \(
    M1_t
    \)
    represent the money supply at time $t$, including currency in circulation and demand deposits, measuring the most liquid components of the money supply.

    \item M2 Money Supply:
    Let
    \(
    M2_t
    \)
    represent the money supply at time $t$, including $M1_t$ plus savings deposits and other near-money assets, capturing broader monetary aggregates.
    \item CPI year-over-year growth rate:
    $
    \pi_t^{\mathrm{YoY}} = \frac{\mathrm{CPI}_t - \mathrm{CPI}_{t-12}}{\mathrm{CPI}_{t-12}},
    $
    measuring long-term inflation pressure.

    \item Purchasing Managers' Index (PMI):
    The official index $\mathrm{PMI}_t$, where $\mathrm{PMI}_t > 50$ indicates expansion and
    $\mathrm{PMI}_t < 50$ indicates contraction.
\end{itemize}

Based on $(\pi_t^{\mathrm{YoY}}, \mathrm{PMI}_t)$, we classify the economy into one of the four
Merrill Lynch regimes $\mathcal{R}_t \in \{\text{Recovery}, \text{Overheating}, \text{Stagflation}, \text{Recession}\}$:
\[
\mathcal{R}_t =
\begin{cases}
\text{Recovery}, & \pi_t^{\mathrm{YoY}} \downarrow, \; \mathrm{PMI}_t > 50, \\
\text{Overheating}, & \pi_t^{\mathrm{YoY}} \uparrow, \; \mathrm{PMI}_t > 50, \\
\text{Stagflation}, & \pi_t^{\mathrm{YoY}} \uparrow, \; \mathrm{PMI}_t < 50, \\
\text{Recession}, & \pi_t^{\mathrm{YoY}} \downarrow, \; \mathrm{PMI}_t < 50. \\
\end{cases}
\]

We use the Purchasing Managers’ Index (PMI) rather than Gross Domestic Product (GDP) as a proxy for economic growth, as PMI is available on a monthly basis, whereas GDP is released only quarterly. This choice ensures consistency with the monthly frequency of the Consumer Price Index (CPI) and enables timely, monthly updates to our allocation decisions.

For each regime $\mathcal{R}_t$, we assign a prior weight vector over industries:
\begin{equation*}
w^{\mathrm{macro}}_{t} = f_{\mathrm{macro}}(\mathcal{R}_t, \Delta M_t),
\qquad
\Delta M_t = \frac{M1_t - M1_{t-1}}{M1_{t-1}} - \frac{M2_t - M2_{t-1}}{M2_{t-1}},
\end{equation*}
    represent the difference in growth rates between $M1_t$ and $M2_t$ at time $t$, reflecting relative liquidity dynamics in the market.
The function $f_{\mathrm{macro}}(\cdot)$ encodes economic intuition from human expertise, including:
\begin{itemize}
    \item Recovery: overweight cyclical sectors (technology, industrials, consumer discretionary).
    \item Overheating: overweight commodities, energy, materials.
    \item Stagflation: overweight defensive sectors (utilities, staples, healthcare).
    \item Recession: overweight bonds or defensive equities (staples, healthcare).
\end{itemize}

For industry $j$, we compute multi-horizon momentum as
\[
M_{j,t} = \sum_{n \in \mathcal{N}} w_n \, R_{j,t}^{(n)},
\quad R_{j,t}^{(n)} = \frac{P_{j,t} - P_{j,t-n}}{P_{j,t-n}},
\]
where $P_{j,t}$ is the index level of industry $j$, and $\mathcal{N}$ is a set of look-back windows.
We then rank industries by $M_{j,t}$ for each horizon $n$, select the top-$m$ industries, and
assign them equal weights:
\[
w^{\mathrm{mom}}_{j,t} =
\begin{cases}
\frac{1}{m}, & j \in \arg\top_m \{ M_{j,t} \}_{j=1}^J, \\
0, & \text{otherwise}.
\end{cases}
\]

The final allocation combines macro-cycle weights and momentum-based weights as:
\[
w^{\mathrm{industry}}_{j,t} = \lambda \, w^{\mathrm{macro}}_{j,t} + (1-\lambda) \, w^{\mathrm{mom}}_{j,t},
\]
where $\lambda \in [0,1]$ controls the relative emphasis on macro regime alignment versus industry momentum.

The reduced stock pool is then defined as
\[
\mathcal{S}_t^{\mathrm{ind}} = \bigl\{\, i \in \mathcal{S}_t^{\mathrm{CSI}}
\;\big|\; \text{industry}(i) \in \arg\top_m w^{\mathrm{final}}_{j,t} \,\}.
\]

\subsection{Firm-Level Multi-Agent Scoring}
Within the reduced stock pool \(\mathcal{S}_t^{\mathrm{ind}}\), we evaluate individual stocks using a set of four specialized agents, \(\mathcal{A} = \{\mathrm{Fundamental}, \mathrm{Technical}, \mathrm{News}, \mathrm{Report}\}\). These agents leverage firm-level features to assess the investment potential of each stock and construct an optimized portfolio. For each agent \(a \in \mathcal{A}\), let \(X^{(a)} \in \mathbb{R}^{n \times T \times d_a}\) denote the panel of features, and \(f_{\phi_a}\) the scoring function. Each stock \(i\) receives an agent-specific score:
\begin{equation*}
z^{(a)}_{t,i} = f_{\phi_a}(X^{(a)}_{i,t-n:t}), \qquad i \in \mathcal{S}_t^{\mathrm{ind}}.
\end{equation*}
where $X^{(a)}_{i,t-n:t}$ represents the panel of features for asset $i$ over the time window from $t-n$ to $t$. The temporal horizon $n$ varies by agent type: the Fundamental agent requires extensive financial data spanning up to five years to capture long-term trends, while the News agent utilizes a shorter horizon of one month to reflect recent sentiment dynamics.

We then collect all agent scores into
\begin{equation*}
z_{t,i} = \big[z^{(a)}_{t,i}\big]_{a\in\mathcal{A}} \in \mathbb{R}^{|\mathcal{A}|}.
\end{equation*}

\subsubsection{Fundamental Agent}
The agent analyzes accounting and financial statement data to evaluate the overall financial health, operating efficiency, management quality, intrinsic value, and innovation-driven growth potential of firms. Key variables include return on equity (ROE), net profit, revenue, and the asset-to-debt ratio.
By integrating these metrics, the Fundamental agent provides a score reflecting long-term value creation and the sustainability of firms.

\subsubsection{Technical Agent}
This agent evaluates stocks based on their historical price and volume data to identify strong technical signals.
It synthesizes dynamic features across multiple temporal horizons, thereby capturing short-term, medium-term, and cross-sectional price dynamics. The indicators include:
\begin{enumerate}[label=(\roman*)]
    \item Momentum: Utilizes sustained price trends via indicators like the Exponential Moving Average (EMA).
    \item Mean Reversion: Identifies price deviations using the Relative Strength Index (RSI).
    \item Volatility: Uses volatility measures such as the Average True Range (ATR).
    \item Statistical Arbitrage: Applies quantitative indicators, e.g., Bollinger Bands, Average Directional Index (ADX), and Hurst exponent.
\end{enumerate}
Through a weighted ensemble of these signals, the Technical agent generates a composite score reflecting the relative trading attractiveness of each stock.

\subsubsection{News Agent}
The News agent functions as a sentiment analyst, focusing on media and news feeds. It analyzes recent news articles for each stock to gauge sentiment intensity. It fetches news data from Baidu and Google and uses the LLM to summarize sentiment. The agent's reasoning includes identifying dominant news themes for the stock and any signals arising from news. By providing the LLM with an updated working memory that includes current volatility, it generates a sentiment polarity score \(s^{\mathrm{news}}_{t,i}\), which quantifies market perception:
\begin{equation*}
z^{\mathrm{news}}_{t,i} = f_{\mathrm{LLM}}(\text{News Articles}_{i,1:t}).
\end{equation*}
The agent evaluates whether positive or negative media narratives are likely to affect near-term stock performance.

\subsubsection{Report Agent}
This agent acts as a specialist in fundamental disclosures and analyst reports. It analyzes the reports from five areas: Investor Research and Inquiry Records, Legal Enforcement and Dishonesty Details, Company Financial Performance Analysis, Stock Distribution Plans and Institutional Investor Holding Proportions for each company. From these, it computes composite scores $z^{\mathrm{report}}$ including the analyst interest level, integrity risk, sentiment score of company management, dividend policy quality, and institutional confidence.:
\begin{equation*}z^{\mathrm{report}}_{t,i} = f_{\mathrm{LLM}}(\text{Reports}_{i,1:t}),
\end{equation*}
thereby capturing institutional sentiment and professional investor expectations.

\subsection{Portfolio Agent: Reinforcement Learning for Portfolio Management}

We utilize RL to integrate signals from fundamental, technical, news, and reports agents for portfolio optimization. By allocating weights across these agents, the RL aims to maximize expected returns. It incorporates state features, such as performance rankings and return histories, and optimizes actions by simulating diverse potential allocations. The agent’s policy is refined through Proximal Policy Optimization (PPO) with action simulation and behavioral cloning, enhancing both stability and convergence in volatile markets.

The final portfolio $\mathcal{P}_t$ is constructed by reweighting the scores from each agent for each stock into scores:
\begin{equation*}
\mathcal{P}t
=
\Big\{{ \rho(z_{t,i}, \mathbf{w}^{industry}_t, \mathbf{w}^{agent}_t) : i \in \mathcal{S}_t^{\mathrm{ind}} }\Big\}
\end{equation*}
where $\rho(z_{t,i}, \mathbf{w}^{industry}_t, \mathbf{w}^{agent}_t)$ represents the aggregated score of stock $i$ obtained using the industry and agent weights $\mathbf{w}^{industry}_t$, $\mathbf{w}^{agent}_t$. The generation of agent weight is formalized as an RL problem. At each trading day $t$, the RL allocates weights of the fundamental, technical, news-sentiment and report analysis LLM agents. The environment evolves as
$
s_{t+1} \sim \mathcal{P}(\cdot \mid s_t, \mathbf{w}^{agent}_t),
$
Driven by next-day market data. The objective is to identify the subset of agents with the portfolio with the highest return:
\begin{equation*}
\pi^*_\theta = \arg\max_{\pi_\theta}\left[\sum_{t=1}^\infty \gamma^{t-1} r_t(s_t, \pi_\theta(s_t)) \right] = \arg\max_{\pi_\theta} Q\left(s_t, \pi_\theta(s_t) \right)
\end{equation*}
with discount factor $\gamma$. Without considering the cost of changing the portfolio, the objective can be viewed as maximizing the return of the next trading day, so the discount factor is set to be close to zero. Considering the noisy reward function and nonstationary environment, a critic that learns the expected immediate reward is proposed to improve the efficiency of the policy gradient.

\textbf{State} $s_t = \big[x_{t}^{(a)}: a \in \mathcal{A}\big] \in \mathcal{S}$,
where each $x^{a \in \mathcal{A}}_t$ encodes the trading returns of $N$ trading days $r_{i,t-1:t-N}$, similar to \cite{ye2020reinforcement}, we augmented the state for RL to select the top-k agents, the relative frequency of agent $i$ being top-k over a sliding window, last action weight $w_{t-1}^{(a)}$ and a reference weighting decision $w_t^{(a)\; ref}$.
$w_t^{(a)\; ref}$ is formed as the allocation weights across the agents on a rolling basis and determined using K-means clustering. Following modern portfolio theory, the weights
are optimized to maximize the expected Sharpe ratio of the resulting portfolio:
\begin{equation*}
\mathbf{w}_t^{ref\star} = \arg\max_{\mathbf{w}^{ref}_t}\;
\frac{\mathbb{E}_{\tau \leq t}[R_\tau(\mathbf{w}^{ref}_t, Y^{(k)})]-r_f}
{\sqrt{\mathrm{Var}_{\tau \leq t}[R_\tau(\mathbf{w}^{ref}_t, Y^{(k)})]}},
\end{equation*}
where \(r_f\) denote the (constant) risk-free rate and $R_\tau(\mathbf{w}^{ref}_t,Y^{(k)})$ denotes the realized return of the portfolio at time $\tau$ formed by
combining the four-agent signals with weights $\mathbf{w}^{ref}_t$.

\textbf{Action} $a_t \in \mathcal{A}$ corresponds to portfolio weights across agents and is generated as a pair of weight vectors:
\begin{equation*}
\tilde{\mathbf{w}}_t^{agent} = \mathrm{softmax}(\pi_{\theta}(s_t)/\tau)
\qquad
\mathbf{w}_t^{agent} = \mathrm{Normalize}\!\big(\operatorname{TopK}(\tilde{\mathbf{w}}_t^{agent} +\beta_{ref}\mathbf{w}_t^{ref\star}) \big)
\end{equation*}
Where $\pi_{\theta}$ the decision policy incorporated with temperature parameter $\tau$, $\tilde{\mathbf{w}}_t^{agent}$ is the raw softmax distribution, while $\mathbf{w}_t^{agent}$ is the masked Top-$k$ allocation actually deployed in the portfolio. Both satisfy $\sum_{a\in \mathcal{A}} w^{(a)}_{t}=1$.

\textbf{Reward} $r_t\in\mathbb{R}^N$ denote the agent rewards on day $t$:
\begin{equation*}
r_{t,i} \;=\;
\lambda_1 (R_t(\mathbf{w}_t^{agent}) - R_t^{\text{def}})
\end{equation*}
with non-negative $\lambda_1$. The term respectively measures excess performance over equal-weight baseline $R_t(\mathbf{w}_t^{agent}) - R_t^{\text{def}}$.

\subsubsection{Action Simulation and Behavior Cloning}

In online learning, where only one state-action pair is updated each trading day, the available samples for training are limited. Inspired by \cite{yang2020deep, zhuang2023behavior}, we incorporate action simulation by generating decisions from a mixture of strategies, thereby increasing the sample size. To enhance the sampling efficiency of the Proximal Policy Optimization (PPO) algorithm using off-policy data, we apply behavioral cloning (BC) to minimize the discrepancy between the behavior policy and the current policy.

\emph{Action simulation} explores the action space by generating multiple potential actions for each trading day. These strategies are designed to explore a diverse set of potential allocations: expert weight, specific weight, uniformly sampled weight and current weight. These four types of action distributions are the expert actions from the previous days, a constant weight from the previous observations leading to the highest return, the weight sampled from a uniform distribution for exploring the action space and the weight from the current policy respectively. Candidate actions are evaluated in the environment, and their outcomes are stored in a replay buffer.

\emph{Behavioral cloning} is a supervised learning technique used to encourage the agent to mimic expert actions during training \cite{shafiullah2022behavior, florence2022implicit}. The expert action is generated from the return by rank the top-k agent returns and reweighted with the sum of one. The BC loss combines mean-squared error (MSE) and cross-entropy (CE):
\begin{equation*}
L_{\text{BC}}=\underbrace{\mathrm{MSE}\big(\mathbf{w}^{\text{pred}}_{\text{mask}},\mathbf{w}^{\text{expert}}_{\text{mask}}\big)}_{\text{matching sparse allocations}}
+\tfrac{1}{2}\underbrace{\mathrm{CE}\big(\mathbf{w}^{\text{expert}}_{\text{mask}},\mathbf{w}^{\text{pred}}_{\text{mask}}\big)}_{\text{target-as-weights cross-entropy}}
\end{equation*}

\begin{algorithm}[th]
\caption{Single-Actor PPO with Critic for Portfolio Allocation}
\begin{algorithmic}[1]
\Require $\theta,\phi$; targets $\theta'\!\gets\!\theta$, $\phi'\!\gets\!\phi$; $\gamma$, $\tau$; $\epsilon$; $E_{\text{PPO}}$; Top-$k$
\Require $\beta_{\text{PPO}}$, $\beta_{\text{MSE}}$, $\beta_{\text{Entropy}}$, $\beta_{\text{BC}}$, $\beta_{\text{decay}}$
\Require $\lambda_1$; $(\alpha_{\text{expert}}, \alpha_{\text{specific}}, \alpha_{\text{uniform}}, \alpha_{\text{current}})$
\State Initialize device-side buffer for $(s,\tilde{\mathbf{w}}^{agent},\mathbf{w}^{agent},r,s',\log\pi_{\text{old}})$
\For{each trading day $t$}
  \State Observe $s_t = \big[x_{t}^{(a)}: a \in \mathcal{A}\big]$
  \State \textbf{Actor:} $\mathbf{w}_t^{agent} = \mathrm{Normalize}\!\big(\operatorname{TopK}(\tilde{\mathbf{w}}_t^{agent} +\beta_{ref}\mathbf{w}_t^{ref\star}) \big)$
  \State Observe rewards $r_{t,i} = \lambda_1 R_t(\mathbf{w}_t^{agent})$, next state $s_{t+1}$
  \State \textbf{Simulate:} $(s_t, \mathbf{w}^{(m)}, r_t^{(m)}, s^{(m)}{t+1})$ using ratios $(\alpha_{\text{expert}}, \alpha_{\text{specific}}, \alpha_{\text{uniform}}, \alpha_{\text{current}})$ with added noise $\mathcal{N}(0,\sigma^2)$
    \For{$b=1:\lfloor\text{buffer}/\text{batch}\rfloor$}
      \State $L_{\text{critic}}(\phi) = \mathbb{E}\left[\big(r + \gamma Q_{\phi'}(s_{t+1}, a_{t+1}) - Q_\phi(s_t, a_t)\big)^2\right]$
        \State $L_{\text{PPO}}(\theta)=-\mathbb{E}\left[\min \big(\rho_t A_t,\mathrm{clip}(\rho_t,1-\epsilon,1+\epsilon )A_t \big) \right]$
        \State $L_{\text{actor}}=\beta_{\text{PPO}}L_{\text{PPO}}+\beta_{\text{MSE}}\,\mathrm{MSE} \big(\mathbf{w}^{\text{pred}}_{\text{mask}},\mathbf{w}^{\text{batch}}_{\text{mask}} \big)+\beta_{\text{Entropy}}\,H(\tilde{\mathbf{w}})+\beta_{\text{BC}}L_{\text{BC}}$
        \Statex \quad Update $\theta \leftarrow \theta - \eta_{\text{actor}}\nabla_\theta L_{\text{actor}}$
    \EndFor
  \State \textbf{Target updates:} $\theta' \leftarrow \tau\theta+(1-\tau)\theta'$, $\phi' \leftarrow \tau\phi+(1-\tau)\phi'$
\EndFor
\end{algorithmic}
\end{algorithm}

\subsubsection{Value Function and Policy Gradient}

The training process is built around the actor–critic paradigm with  Proximal Policy Optimization (PPO).

The \emph{critic} is trained to predict the value function \(Q_{\phi}(s_t, a_t)\), which represents expected returns from state \(s_t\). It is optimized by minimizing:
\begin{equation*}
L_{\text{critic}}(\phi) = \mathbb{E}_t \left[ \left( r_t+\gamma Q_\phi\left(s_{t+1}, a_{t+1}\right)-Q_\phi\left(s_t, a_t\right) \right)^2 \right].
\end{equation*}

The \emph{actor} updates its policy \(\pi_\theta(a|s)\) using the PPO clipped surrogate loss:
\begin{equation*}
L_{\text{PPO}}(\theta) = \mathbb{E}_t \left[\min\left(\rho_t A_t, \; \mathrm{clip}(\rho_t, 1-\epsilon, 1+\epsilon) A_t \right)\right],
\end{equation*}
where: \(\rho_t = \frac{\pi_\theta(a_t|s_t)}{\pi_{\theta_{\text{old}}}(a_t|s_t)}\) is the probability ratio between the new and old policies,
\(A_t\):
\begin{equation*}
A_t = r_t + \gamma Q_{\phi}(s_{t+1}, a_{t+1}) - Q_{\phi}(s_t, a_t),
\end{equation*}
where \(V_{\phi}(s_t)\) is the value function learned by the critic, and \(r_t\) is the reward at time \(t\).

The actor’s total loss integrates PPO, imitation, and entropy regularization:
\begin{equation*}  L_{\text{actor}}=\beta_{PPO}L_{\text{PPO}}+\beta_{MSE}\,\mathrm{MSE}(\mathbf{w}^{\text{pred}}_{\text{mask}},\mathbf{w}^{\text{batch}}_{\text{mask}})+\beta_{Entropy}\,H(\tilde{\mathbf{w}})+\beta_{\!BC}L_{\text{BC}}
\end{equation*}

Performance is evaluated out-of-sample by applying the learned policy to the next day’s returns, typically using closing prices. After the market is closed, the agent is trained using the updated data and decides the weight allocation for the next day. This ensures the model is trained on historical information while tested on unseen market dynamics.

\subsection{Risk Control Agent}
To further investigate the risk-neutral performance of our model, we design a risk scaling agent. Specifically, IF contracts can be traded on the China Financial Futures Exchange as a hedging instrument. This agent manages portfolio exposure by dynamically adjusting positions in response to market volatility. It implements a risk scaling algorithm, which reduces positions when market volatility is extreme and gradually increases them when volatility is moderate (up to a maximum scale of 1).

Specifically, after portfolio construction, the weight of the assets can be generated:
\begin{equation*}
    \mathbf{p}_t \;=\; f_\theta\!\big(\,\mathbf{F}_{:,1:t,:},\, \mathbf{S}_{1:t,:},\, \mathbf{B}_{1:t,:}\big), \qquad \mathbf{p}_t \in \mathbb{R}^{N},
\qquad
\mathbf{1}^\top \mathbf{p}_t = 1,
\qquad
\mathbf{p}_t \succeq \mathbf{0},
\label{eq:weights}
\end{equation*}
where \(f_\theta(\cdot)\) is a parameterized allocation rule that maps information up to time \(t\) to weights and \(\theta\) denotes learnable parameters and the nonnegativity enforces long-only allocations.

The Risk Control agent scales exposure dynamically to stabilize portfolio volatility around a pre-specified target level $\sigma_{\mathrm{tgt}}$. Following \cite{zhang2019deepreinforcementlearningtrading}, the scaling factor is defined as
\begin{equation*}
\beta_t = \frac{\sigma_{\mathrm{tgt}}}{\widehat{\sigma}_{t-1}}, \qquad
\widehat{\sigma}_{t-1}^2 = (1-\lambda)\sum_{j=1}^{n}\lambda^{\,j-1} r_{t-j}^2,
\qquad 0<\lambda<1,
\end{equation*}
where $\widehat{\sigma}_{t-1}$ is the estimated market volatility at time $t-1$.
The volatility estimate is computed using an exponentially weighted moving standard deviation of past
returns $r_t$ over an $n$-day window with $\lambda$ being the decay factor.

The final tradable portfolio weights are therefore given by
\begin{equation*}
\tilde{\mathbf{p}}_t = \beta_t \, \mathbf{p}_t = \beta_t \, f_\theta\!\big(\,\mathbf{F}_{:,1:t,:},\, \mathbf{S}_{1:t,:},\, \mathbf{B}_{1:t,:}\big)
\end{equation*}
providing a volatility-targeting adjustment, ensuring that position sizes are adaptively scaled—expanded (up to full allocation) in tranquil market conditions and contracted under heightened volatility—thereby improving Sharpe ratio and Calmar ratio.

\section{Experiments}\label{sec:experiment}
This section delineates the experimental framework used to assess the performance of our multi-agent trading system. It provides an exposition of the baseline models, LLM configurations, and evaluation metrics utilized to ensure a robust and systematic analysis of the system's efficacy.

\subsection{Experimental Setup}
We use Qwen3-32B as the foundational LLM as agents' reasoning core. Recognizing that pretrained models may contain implicit knowledge of financial markets, we obfuscate industry \& company identifiers before submission to the model in order to prevent information leakage. To prevent look-ahead bias, agents’ decisions on trading day $T$ are generated exclusively from masked inputs observed up to day $T-1$. The outputs are aggregated by a portfolio manager, who proposes the preliminary asset allocation. This allocation is then processed by a dedicated risk manager agent, which incorporates recent volatility dynamics to adjust exposures, thereby aligning the portfolio with risk-control objectives.

\paragraph{Training and testing protocol.}
We assess the proposed framework through backtesting simulation conducted over the period from January 1, 2019, to December 31, 2024, using the CSI 300 constituent stock pool. We split the data into training and testing periods: from 2019-01-01 to 2023-12-31 as the training period and from 2024-01-01 to 2024-12-31 as the testing period.

In the training period, parameter search is conducted for each agent independently. For each firm-level agent (fundamental, technical, report, and news), we design a time-series optimizer that leverages arbitrage principles to produce daily trading signals, compensating for the low update frequency of financial, news, and report data. Parameter search is carried out within each optimizer. For agent $a$,
parameters $\phi_a$ are optimized to maximize its standalone annualized Sharpe ratio:
\begin{equation}
\phi_a^\star = \arg\max_{\phi_a}\;
\frac{\mathbb{E}_t\!\left[R^{(a)}_t(\phi_a, Y^{(k)})\right]-r_f}
{\sqrt{\mathrm{Var}_t\!\left[R^{(a)}_t(\phi_a, Y^{(k)})\right]}},
\label{eq:agent-sharpe}
\end{equation}
where $R^{(a)}_t$ is the realized portfolio return constructed solely from agent $a$’s scores.
The optimized parameters $\phi_a^\star$ are then fixed and directly applied to the testing period
without re-estimation.

\subsection{Benchmark}
To assess the efficacy of our system, we conduct  analysis against a range of established baselines across multiple categories. These include conventional proxy indicators—specifically, MACD \citep{lim2020enhancingtimeseriesmomentum}, KDJ, RSI, sign(R) \citep{moskowitz2012}, and SMA—as well as the MASS framework \citep{guo2025mass}. Below we introduce a detailed list of benchmarks:
\begin{enumerate}[label=(\roman*)]
    \item MASS (Multi-Agent Simulation Scaling for Portfolio Construction): MASS \citep{guo2025mass} is a framework that leverages large-scale collaborative agents to systematically construct and optimize investment portfolios.
    \item {CSI 300}: The CSI 300 Index is a capitalization-weighted stock market index designed to reflect the performance of the top 300 stocks listed on the Shanghai and Shenzhen stock exchanges. It covers a diverse range of industries and serves as a key benchmark for the Chinese A-share market.
    \item {SMA (Simple Moving Average)}: This trend-following strategy generates signals from crossovers of a 5-period short-term SMA and a 10-period long-term SMA.

    \item {RSI (Relative Strength Index)}: RSI measures price momentum over a 10-period look-back, entering long positions when RSI \textless~30 (oversold) and closing when RSI \textgreater~70 (overbought).

    \item {Sign}: This strategy focus only on the sign of the past 20 days return, triggering buy signals when sign(20) \textgreater~0 and sell signals when sign(20) \textless~0.

    \item {KDJ}: This stochastic oscillator uses a 9-period look-back to generate trading signals based on K and D line crossovers, with the J line confirming momentum.

    \item {MACD (Moving Average Convergence Divergence)}: This trend-following strategy uses a fast EMA (12 periods), slow EMA (26 periods), and signal line (9 periods) to generate buy/sell signals based on MACD-signal line crossovers.
\end{enumerate}

System performance is quantified using eight key metrics: Cumulative Return (CR), Annualized Return (AR), Annualized Standard Deviation (STD), Downside Deviation (DD) \citep{ang2006}, Sharpe Ratio \citep{sharpe1994}, Sortino Ratio \citep{rollinger2013}, Maximum Drawdown (MDD), and Calmar Ratio \citep{young1991}. The precise formulations for these metrics are detailed in the e-companion.

\begin{figure}[htbp!]
\centering
\caption{Backtesting Results Compared with Baseline Strategies and the CSI 300 Index (Training Sample) (a) Cumulative Returns, (b) Excess Returns Relative to the CSI 300 Index}
\begin{subfigure}[b]{\textwidth}
         \centering
\includegraphics[width=0.7\textwidth]{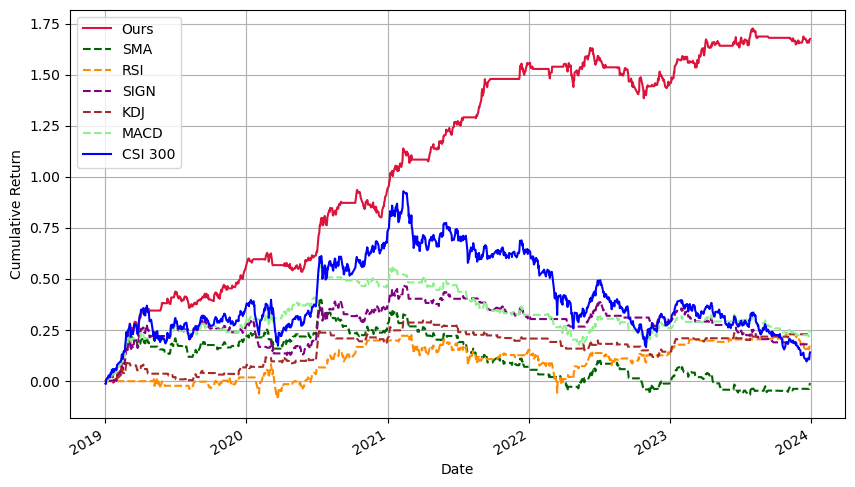}
\vspace{-5mm}
\caption{\label{fig:Long_only_training}}
\end{subfigure}
\begin{subfigure}[b]{\textwidth}
         \centering
\includegraphics[width=0.7\textwidth]{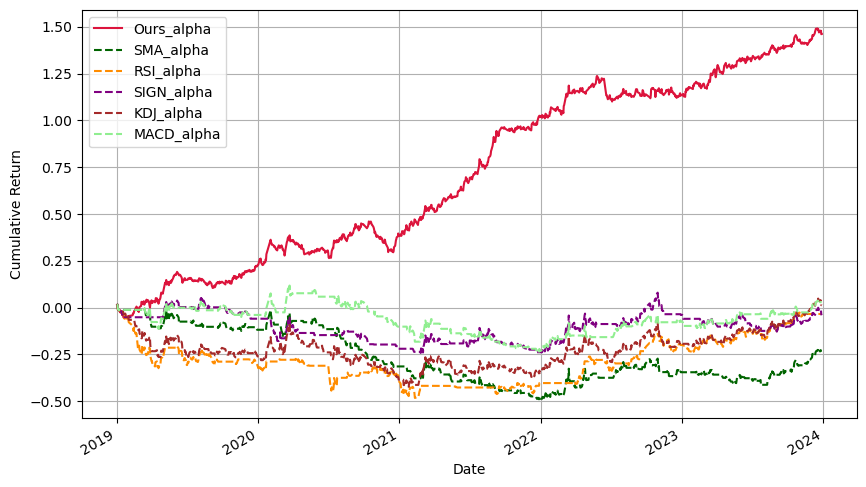}
\vspace{-5mm}
\caption{\label{fig:Alpha_training}}
\end{subfigure}

\end{figure}

\begin{table}[ht]
\TABLE{Performance in Training Sample
\label{tab:Long_only_training_table}
}
{
\resizebox{\textwidth}{!}{%
\begin{tabular}{l*{8}{r}}
\toprule
\textbf{Strategy} & \textbf{CR (\%) $\uparrow$} & \textbf{AR (\%) $\uparrow$} & \textbf{STD (\%) $\uparrow$} & \textbf{DD (\%) $\uparrow$} & \textbf{Sharpe $\uparrow$} & \textbf{Sortino $\uparrow$} & \textbf{MDD (\%) $\downarrow$} & \textbf{Calmar $\uparrow$} \\
\midrule
SMA             & -1.06   & -0.21   & 13.05   & 9.03   & -0.02  & -0.02  & -46.50   & -0.01 \\
RSI             & 18.61   & 3.72    & 13.41   & 9.61   & 0.28   & 0.39   & -28.48   & 0.13  \\
SIGN            & 17.93   & 3.59    & 13.39   & 9.31   & 0.27   & 0.39   & -28.82   & 0.12  \\
KDJ             & 23.33   & 4.67    & \textbf{8.23}    & \textbf{4.77}    & 0.57   & 0.98   & \textbf{-17.89}   & 0.26  \\
MACD            & 24.93   & 4.99    & 12.97   & 8.61   & 0.38   & 0.58   & -39.41   & 0.13  \\
\textbf{Ours} & \textbf{167.52} & \textbf{34.77} & 18.94   & 11.41   & \textbf{1.84} & \textbf{3.05} & -24.58   & \textbf{1.42} \\
\bottomrule
\end{tabular}}
}
{}
\end{table}

\begin{table}[ht]
\TABLE{Performance of Excess Returns in Training Sample \label{tab:Alpha_training_table}}
{
\resizebox{\textwidth}{!}{%
\begin{tabular}{l*{8}{r}}
\toprule
\textbf{Strategy} & \textbf{CR (\%) $\uparrow$} & \textbf{AR (\%) $\uparrow$} & \textbf{STD (\%) $\uparrow$} & \textbf{DD (\%) $\downarrow$} & \textbf{Sharpe $\uparrow$} & \textbf{Sortino $\uparrow$} & \textbf{MDD (\%) $\uparrow$} & \textbf{Calmar $\uparrow$} \\
\midrule
SMA\_alpha        & -23.11  & -4.62   & 14.20   & 9.85   & -0.33  & -0.47  & -50.55   & -0.09 \\
RSI\_alpha        & -3.44   & -0.69   & \textbf{13.86}   & 9.91   & -0.05  & -0.07  & -49.69   & -0.01 \\
SIGN\_alpha       & -4.12   & -0.82   & 13.88   & 9.64   & -0.06  & -0.09  & -29.35   & -0.03 \\
KDJ\_alpha        & 1.28    & 0.26    & 17.44   & 11.86  & 0.01   & 0.02   & -43.87   & 0.01  \\
MACD\_alpha       & 2.88    & 0.58    & 14.28   & 9.57   & 0.04   & 0.06   & -35.31   & 0.02  \\
\textbf{Ours\_alpha}   & \textbf{146.10} & \textbf{30.33} & 16.57   & \textbf{10.07}   & \textbf{1.83} & \textbf{3.01} & \textbf{-16.49}   & \textbf{1.84} \\
\bottomrule
\end{tabular}}
}
{}
\end{table}

\begin{figure}[htbp!]
\centering
\caption{Backtesting Results Compared with Baseline Strategies and the CSI 300 Index (Testing Sample) (a) Cumulative Returns, (b) Excess Returns Relative to the CSI 300 Index}
\begin{subfigure}[b]{\textwidth}
         \centering
\includegraphics[width=0.7\textwidth]{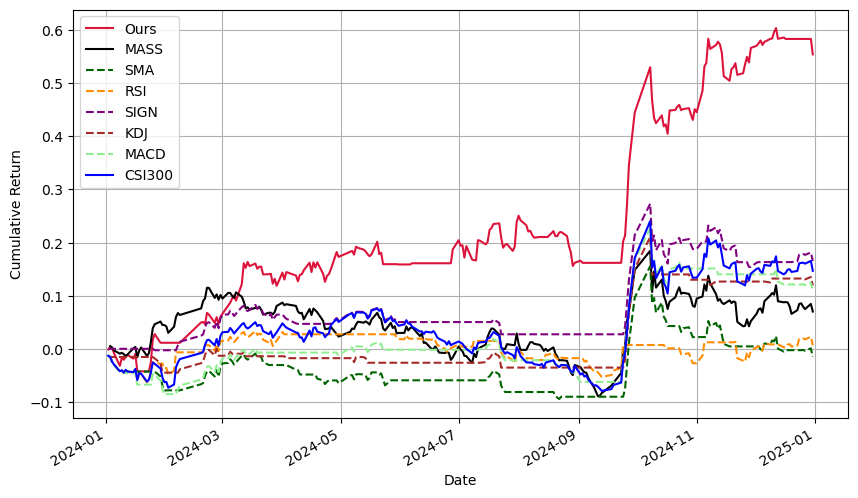}
\vspace{-5mm}
\caption{\label{fig:Long_only_testing}}
\end{subfigure}
\begin{subfigure}[b]{\textwidth}
         \centering
\includegraphics[width=0.7\textwidth]{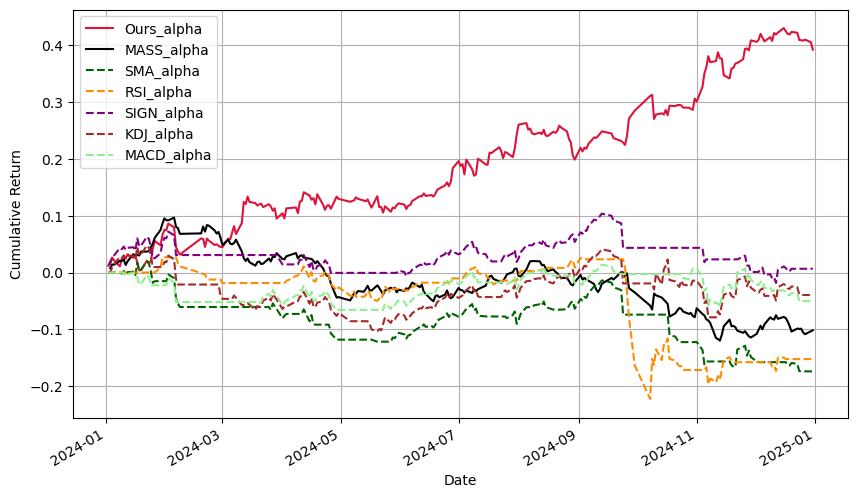}
\vspace{-5mm}
\caption{\label{fig:Alpha_testing}}
\end{subfigure}

\end{figure}

\begin{table}[ht!]
\TABLE{Performance in Testing Sample \label{tab:Long_only_testing_table}
}
{\resizebox{\textwidth}{!}{%
\begin{tabular}{l*{8}{r}}
\toprule
\textbf{Strategy} & \textbf{CR (\%) $\uparrow$} & \textbf{AR (\%) $\uparrow$} & \textbf{STD (\%) $\uparrow$} & \textbf{DD (\%) $\uparrow$} & \textbf{Sharpe $\uparrow$} & \textbf{Sortino $\uparrow$} & \textbf{MDD (\%) $\downarrow$} & \textbf{Calmar $\uparrow$} \\
\midrule
SMA             & -1.53   & -1.53   & 17.68   & 10.95   & -0.09 & -0.14 & -17.09   & -0.09 \\
RSI             & 0.64    & 0.64    & \textbf{11.80}   & \textbf{7.42}    & 0.05  & 0.09  & \textbf{-8.64}    & 0.07  \\
SIGN            & 16.53   & 16.53   & 17.98   & 10.66   & 0.92  & 1.55  & -11.89   & 1.39  \\
KDJ             & 11.91   & 11.91   & 15.34   & 8.54    & 0.78  & 1.40  & -9.14    & 1.30  \\
MACD            & 10.84   & 10.84   & 18.19   & 10.66   & 0.60  & 1.02  & -11.84   & 0.92  \\
MASS            & 7.01    & 7.01    & 21.79   & 14.57   & 0.32  & 0.48  & -20.41   & 0.34  \\
\textbf{Ours} & \textbf{55.41}   & \textbf{55.41}   & 28.20   & 14.90   & \textbf{1.96}   & \textbf{3.72}   & -12.52   & \textbf{4.43} \\
\bottomrule
\end{tabular}}
}
{}
\end{table}

\begin{table}[ht!]
\TABLE{Performance of Excess Returns in Testing Sample
\label{tab:Alpha_testing_table} }
{\resizebox{\textwidth}{!}{%
\begin{tabular}{lrrrrrrrr}
\toprule
\textbf{Strategy} & \textbf{CR (\%) $\uparrow$} & \textbf{AR (\%) $\uparrow$} & \textbf{STD (\%) $\downarrow$} & \textbf{DD (\%) $\downarrow$} & \textbf{Sharpe $\uparrow$} & \textbf{Sortino $\uparrow$} & \textbf{MDD (\%) $\uparrow$} & \textbf{Calmar $\uparrow$} \\
\midrule
SMA\_alpha         & -17.40 & -17.40 & 11.94  & 10.09  & -1.46 & -1.72 & -19.41 & -0.90 \\
RSI\_alpha         & -15.23 & -15.23 & 17.79  & 14.51  & -0.86 & -1.05 & -24.94 & -0.61 \\
SIGN\_alpha        & 0.66   & 0.66   & 11.49  & 9.15   & 0.06  & 0.07  & -12.20 & 0.05  \\
KDJ\_alpha         & -3.96  & -3.96  & 14.85  & 11.48  & -0.27 & -0.35 & -15.12 & -0.26 \\
MACD\_alpha        & -5.03  & -5.03  & \textbf{11.18}  & \textbf{8.77}   & -0.45 & -0.57 & -7.43  & -0.68 \\
MASS\_alpha        & -10.16 & -10.16 & 12.04  & 9.11   & -0.84 & -1.12 & -21.66 & -0.47 \\
\textbf{Ours\_alpha}
                   & \textbf{39.22}  & \textbf{39.22}  & 17.60  & 10.56   & \textbf{2.23}  & \textbf{3.71}  & \textbf{-6.44}  & \textbf{6.09}  \\
\bottomrule
\end{tabular}%
}}
{}
\end{table}

\subsection{Main Results}

In contrast to existing multi-agent frameworks, such as MASS, which relies on rule-based optimization without incorporating risk scaling, our proposed system exhibits superior performance.

Figure~\ref{fig:Long_only_training} and Table~\ref{tab:Long_only_training_table} show that our framework consistently outperforms benchmark models in the training period. It achieves a cumulative return (CR) of 167.52\%, a Sharpe ratio (SR) of 1.84, a Sortino ratio of 3.05, and a Calmar ratio of 1.42. For excess returns, the system records a cumulative return of 146.10\%, with SR, Sortino, and Calmar ratios of 1.83, 3.01, and 1.84, respectively. Compared with both traditional methods (RSI and KDJ) and multi-agent systems such as MASS, ContestTrade demonstrates stronger profitability and improved risk-adjusted performance, confirming the efficacy of its competitive multi-agent design.

The framework also exhibits strong out-of-sample generalization. As presented in Figure~\ref{fig:Long_only_testing} and Table~\ref{tab:Long_only_testing_table}, it yields a cumulative return of 55.41\% in the testing period, outperforming the best baseline by 38.88%

As evidenced in Figure~\ref{fig:Alpha_testing} and Table~\ref{tab:Alpha_testing_table}, our framework demonstrates robust efficacy, consistently achieving enhanced risk-adjusted returns and reduced volatility relative to baseline strategies.

\begin{figure}[htbp]
\FIGURE{\includegraphics[width=0.7\textwidth]{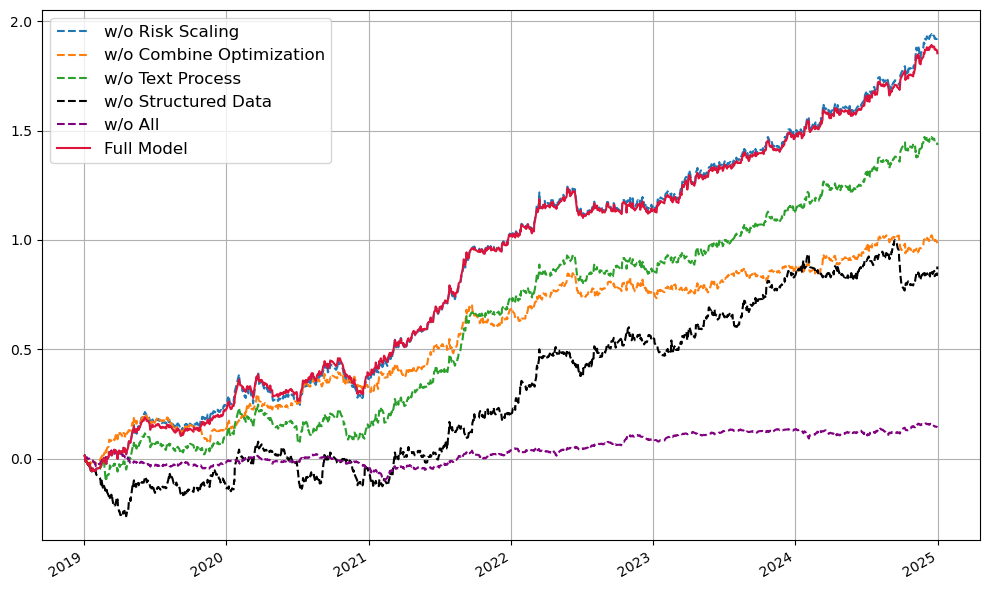}
}
{Comparative Performance of the Full Model and Ablated Configurations Over Time \label{fig:Ablation}}
{``w/o'' indicates removing the specified configuration.}

\end{figure}

\begin{table}[ht]
\TABLE{Ablation Study of Model Performance
\label{tab:ablation_results}}
{
\resizebox{\textwidth}{!}{%
\begin{tabular}{lrrrrrrrr}
\toprule
\textbf{Model} & \textbf{CR (\%) $\uparrow$} & \textbf{AR (\%) $\uparrow$} & \textbf{STD (\%) $\uparrow$} & \textbf{DD (\%) $\uparrow$} & \textbf{Sharpe $\uparrow$} & \textbf{Sortino $\uparrow$} & \textbf{MDD (\%) $\downarrow$} & \textbf{Calmar $\uparrow$} \\
\midrule
\textbf{Full Model}  & 185.33 & 32.08 & 16.75 & 10.15  & \textbf{1.92} & \textbf{3.16} & -16.49 & 1.95 \\
w/o Risk Scaling  & \textbf{190.27} & \textbf{32.93} & 18.05 & 11.10  & 1.82 & 2.97 & -16.49 & \textbf{2.00} \\
w/o Combine Opt.      & 98.56  & 17.06 & 13.72 & 8.94  & 1.24 & 1.91 & \textbf{-13.23} & 1.29 \\
w/o Text Process      & 143.36  & 24.81 & 16.57 & 10.57 & 1.50 & 2.35 & -17.16 & 1.45 \\
w/o Structured Data   & 87.87  & 15.21  & 19.57 & 12.24 & 0.78 & 1.24 & -27.81 & 0.55 \\
w/o All               & 14.36  & 2.49  & \textbf{4.99} & \textbf{3.45} & 0.50 & 0.72 & -13.83 & 0.18 \\
\bottomrule
\end{tabular}%
}
}
{}
\end{table}

\subsection{Ablation Configurations}
To rigorously evaluate the contributions of individual components within our AI-based fundamental investing framework, we conduct an ablation study by systematically disabling key modules.

\begin{enumerate}
[label=(\roman*)]
\item{Without Risk Scaling.}
In this configuration, the risk scaling mechanism is deactivated, resulting in a portfolio without volatility control. This setup isolates the role of risk management in improving the Sortino Ratio and reducing maximum drawdown, highlighting its effectiveness in stabilizing returns.

\item{Without Combined Optimization.}
The combined optimization module is excluded, and final trading signals are equally weighted across all assets. This configuration evaluates the incremental benefits of the optimization process in enhancing portfolio performance metrics, such as risk-adjusted returns.

\item{Without Text Processing (Remove News and Report agents).}
The text-processing modules, including the Report and News agents, are removed, thereby eliminating the framework’s ability to incorporate textual information. This ablation assesses the value of qualitative data in informing investment decisions.

\item{Without Structured Data (Remove Fundamental and Technical agents).}
In this setting, the framework relies exclusively on news and report agents, omitting signals derived from the Fundamental agent and the Technical agent. This configuration quantifies the contribution of structured data to the framework’s predictive accuracy.

\item{Without All Components.}
All key components—risk scaling, combined optimization, text processing, and structured data—are disabled, resulting in a baseline portfolio that equally weights all constituents of the CSI 300 index. This setup serves as a control, illustrating the collective impact of the proposed innovations on portfolio performance.
\end{enumerate}

As evidenced by Figure~\ref{fig:Ablation} and Table~\ref{tab:ablation_results}, the ablation of any individual component substantially impairs the performance of the AI-based fundamental investing framework, with the complete removal of all components resulting in severe degradation of portfolio outcomes. This underscores the critical and synergistic role of each module in achieving optimal results.

\section{Conclusion}\label{sec:conclusion}

In this paper,
we proposed a hierarchical AI multi-agent framework for fundamental equity investing.
The framework organizes the investment process along a macro–industry–firm path and implements it with a coordinated set of agents. It builds equity portfolios from firm fundamentals and runs a systematic trading workflow that combines fundamental signals, technical timing, sentiment inputs, portfolio optimization, and risk scaling.

The framework integrates specialized agents for research, portfolio construction, and risk control, which supports adaptability and robustness in China's volatile A-share market. It delivers a unified analysis across financial, technical, and sentiment dimensions. In empirical tests, the system outperforms standard benchmarks and non-hierarchical multi-agent baselines on key metrics, delivering higher cumulative returns, stronger risk-adjusted performance, and lower downside risk.

Future work includes extending the framework to multiple asset classes and international markets to assess robustness across market regimes. Another direction is adding human-in-the-loop oversight so expert judgment can complement algorithmic decisions.

\bibliographystyle{informs2014} %
\bibliography{references} %

\ECSwitch

\ECHead{Supplementary Material}

This supplementary material provides additional details to support the main document. It includes a high-level illustration of our framework and detailed information on dataset construction.

\section{Performance Metrics}\index{Performance Metrics}
The performance of the portfolio is evaluated using the following metrics:

\begin{enumerate}[label=(\roman*)]
    \item Cumulative Return (CR): The total return of the portfolio over a specific period, reflecting the overall growth of the investment.
    \item Annualized Return (AR): The average return of the portfolio per year, providing a standardized measure for comparing performance across different time frames.
    \item Annualized Standard Deviation (STD): A measure of the portfolio's volatility, indicating the degree of dispersion of returns around the average return.
    \item Downside Deviation (DD): Also known as \emph{downside risk}, this metric measures the volatility of only the negative returns. It focuses on the risk of losses, providing a more specific view of the portfolio's exposure to adverse movements.
    \item Sharpe Ratio: A risk-adjusted measure of return, calculated as the annualized return minus the risk-free rate, divided by the annualized standard deviation. A higher Sharpe Ratio indicates a better return for a given level of risk. The formula is:
    $$ \text{Sharpe Ratio} = \frac{AR - R_f}{STD} $$
    where \( AR \) is the annualized return, \( R_f \) is the risk-free rate, and \( STD \) is the annualized standard deviation of portfolio returns. In this study, we assume the risk-free rate (\( R_f \)) is 0.
    \item Sortino Ratio: A variant of the Sharpe Ratio that uses downside deviation instead of total standard deviation as the risk measure. This ratio assesses the return generated per unit of downside risk. The formula is:
    $$ \text{Sortino Ratio} = \frac{AR - R_f}{DD} $$
    where \( AR \) is the annualized return, \( R_f \) is the risk-free rate, and \( DD \) is the annualized downside deviation of portfolio returns.
    \item Maximum Drawdown (MDD): This metric represents the largest peak-to-trough decline in the portfolio's value over a specified period. It shows the maximum observed loss from any peak, serving as an indicator of capital preservation risk.
    \item Calmar Ratio: A risk-adjusted performance metric that evaluates the annualized excess return relative to the maximum drawdown, defined as:
    $$ \text{Calmar Ratio} = \frac{AR - R_f}{|MDD|} $$
    where \( AR \) is the annualized return, \( R_f \) is the risk-free rate, and \( |MDD| \) is the absolute value of the maximum drawdown.
\end{enumerate}

\section{Dataset Details}

\subsection{Macro-level Data Details}
\begin{longtable}{p{5cm} p{10cm}}
    \toprule
    \textbf{Variable} & \textbf{Definition} \\
    \midrule
    CPI Year-over-Year Growth Rate & Calculated as $\pi_t^{\mathrm{YoY}} = \frac{\mathrm{CPI}_t - \mathrm{CPI}_{t-12}}{\mathrm{CPI}_{t-12}}$, measuring long-term inflation pressure. \\
    M1 Money Supply & Represents the money supply at time $t$, including currency in circulation and demand deposits, measuring the most liquid components of the money supply. \\
    M2 Money Supply & Represents the money supply at time $t$, including M1 plus savings deposits, money market securities, and other near-money assets, capturing broader monetary aggregates. \\
    Purchasing Managers' Index (PMI) & The official index $\mathrm{PMI}_t$, where $\mathrm{PMI}_t > 50$ indicates expansion and $\mathrm{PMI}_t < 50$ indicates contraction. \\
    \bottomrule
\end{longtable}

\vspace{0.5cm} %

\subsection{Industry-level Data Details}
\begin{longtable}{p{5cm} p{10cm}}
    \toprule
    \textbf{Variable} & \textbf{Definition} \\
    \midrule
    Industry Index Return & Daily return rates of industry index based on the industry classification codes of listed companies published by the China Securities Regulatory Commission. \\
    \bottomrule
\end{longtable}

\vspace{0.5cm} %
\clearpage
\subsection{Firm-level Data Details}
\begin{longtable}{p{5cm} p{10cm}}
    \toprule
    \textbf{Variable} & \textbf{Definition} \\
    \midrule
    \endhead
    Capital Expenditures (CapEx) & The money an organization or corporate entity spends to buy, maintain, or improve its fixed assets, such as buildings, vehicles, equipment, or land. \\
    Cash \& Cash Equivalents & Cash on hand and highly liquid short-term investments. \\
    Close & \\
    Current Assets & Assets expected to be converted into cash within one year. \\
    Current Liabilities & Short-term obligations due within one year. \\
    Dividends paid & Cash or stock payments distributed to shareholders or holders of certain equity-based awards. \\
    Earnings Before Interest \& Taxes & Company's operating profit without interest expenses and income taxes. \\
    Earnings before tax & The money retained internally by a company before deducting tax expenses. \\
    Earnings Per Share (EPS) & Portion of a company's profit allocated to each outstanding share of common stock. \\
    Enterprise Value / EBIT (EV/EBIT) & Valuation ratio that compares a company's enterprise value (EV) to its earnings before interest and taxes (EBIT). \\
    Enterprise Value / EBITDA (EV/EBITDA) & Valuation ratio comparing enterprise value to cash earnings (EBITDA). \\
    Free Cash Flow (FCF) & The money that a company has available to repay its creditors or pay dividends and interest to investors. \\
    Goodwill & Excess paid in acquisitions above fair value of net assets. \\
    Gross Profit Margin & Percentage of revenue that exceeds the cost of goods sold. \\
    Intangible Assets & Non-physical assets including intellectual property, patents, and goodwill. \\
    Interest Expense & Cost incurred by an entity for borrowing funds. \\
    MarketValue & Current value of a publicly traded company, based on the total dollar amount that all of its outstanding shares are worth. \\
    Net Income & Remains from a company's total revenues after deducting all operating costs, taxes, interest, and other expenses. \\
    Operating Costs & Daily expenses necessary to maintain, operate, and administer a business. \\
    Operating Margin & Profitability ratio that measures revenue after covering the operating and non-operating expenses. \\
    Operating Profit & Total earnings from a company's core business operations excluding deductions of interest and tax. \\
    Operating Revenue & The money a company generates from its primary business activities. \\
    Paid-in Capital & Capital contributed by shareholders. \\
    R\&D Expenses & The money companies spend on innovation and improving their products, services, technologies, and processes. \\
    Return on Equity (ROE) & Profitability of a business in relation to its equity. \\
    Return on Invested Capital & Profitability ratio measuring return relative to invested capital. \\
    Revenue & The total amount of income generated by the sale of goods and services related to the primary operations of a business. \\
    Shareholders' Equity & Total amount of assets that a company would retain if it paid all of its debts. \\
    Total Assets & Sum of all owned assets (current + non-current). \\
    Total Liabilities & All debts and obligations of a company. \\
    \bottomrule
\end{longtable}

\begin{table}[ht]
\caption{Descriptive statistics of our dataset} \label{tab:descriptive_stats}
\resizebox{\columnwidth}{!}{
    \begin{tabular}{lllllllll}
\toprule
 & count & mean & std & min & 25\% & 50\% & 75\% & max \\
\midrule
CPI YoY Growth Rate & \num{8.40e1} & \num{1.552381e0} & \num{1.351490e0} & \num{-8.00e-1} & \num{3.75e-1} & \num{1.55e0} & \num{2.4725e0} & \num{5.38e0} \\
M1 Money Supply & \num{8.40e1} & \num{6.726956e5} & \num{1.746188e5} & \num{5.17036e5} & \num{5.57053e5} & \num{6.231289e5} & \num{6.744551e5} & \num{1.12012e6} \\
M2 Money Supply & \num{8.40e1} & \num{2.366131e6} & \num{4.464552e5} & \num{1.720814e6} & \num{1.950588e6} & \num{2.307211e6} & \num{2.809994e6} & \num{3.135322e6} \\
Purchasing Managers’ Index (PMI) & \num{8.40e1} & \num{4.9985714e1} & \num{1.886358e0} & \num{3.57e1} & \num{4.94e1} & \num{5.01e1} & \num{5.09e1} & \num{5.26e1} \\
Industry Index Return & \num{1.32179e5} & \num{5.46e-4} & \num{2.1171e-2} & \num{-1.20314e-1} & \num{-9.50e-3} & \num{7.20e-5} & \num{1.0181e-2} & \num{2.675919e0} \\
Capital Expenditures & \num{1.40e4} & \num{2.85e9} & \num{1.29e10} & \num{0.00e0} & \num{1.31e8} & \num{4.71e8} & \num{1.60e9} & \num{3.31e11} \\
Cash \& Cash Equivalents & \num{1.40e4} & \num{2.67e10} & \num{1.90e11} & \num{0.00e0} & \num{7.38e8} & \num{2.12e9} & \num{6.81e9} & \num{4.04e12} \\
Close & \num{2.98e6} & \num{1.78e2} & \num{7.80e2} & \num{1.13e0} & \num{2.05e1} & \num{5.46e1} & \num{1.40e2} & \num{6.58e4} \\
Current Assets & \num{1.31e4} & \num{2.66e10} & \num{9.37e10} & \num{0.00e0} & \num{2.32e9} & \num{6.06e9} & \num{1.67e10} & \num{2.28e12} \\
Current Liabilities & \num{1.31e4} & \num{2.33e10} & \num{8.19e10} & \num{0.00e0} & \num{1.49e9} & \num{4.62e9} & \num{1.44e10} & \num{1.76e12} \\
Dividends paid & \num{1.38e4} & \num{2.31e9} & \num{1.27e10} & \num{0.00e0} & \num{1.13e8} & \num{3.66e8} & \num{1.16e9} & \num{1.13e12} \\
EBIT & \num{1.40e4} & \num{4.98e9} & \num{2.32e10} & \num{-6.52e10} & \num{2.20e8} & \num{8.06e8} & \num{2.41e9} & \num{4.25e11} \\
Earnings before tax & \num{1.40e4} & \num{4.60e9} & \num{2.30e10} & \num{-7.13e10} & \num{1.46e8} & \num{6.63e8} & \num{2.07e9} & \num{4.25e11} \\
EPS & \num{1.40e4} & \num{6.13e-1} & \num{1.58e0} & \num{-1.65e1} & \num{1.10e-1} & \num{3.60e-1} & \num{8.01e-1} & \num{6.86e1} \\
EV/EBIT & \num{1.40e4} & \num{1.55e1} & \num{9.24e2} & \num{-8.85e4} & \num{5.67e0} & \num{1.80e1} & \num{3.49e1} & \num{1.65e4} \\
EV/EBITDA & \num{1.40e4} & \num{2.30e1} & \num{4.23e2} & \num{-2.66e4} & \num{4.93e0} & \num{1.44e1} & \num{2.95e1} & \num{3.25e4} \\
Free Cash Flow (FCF) & \num{1.40e4} & \num{-3.79e10} & \num{2.72e11} & \num{-6.09e12} & \num{-3.13e9} & \num{-5.22e8} & \num{8.65e7} & \num{1.60e11} \\
Goodwill & \num{7.42e3} & \num{9.57e8} & \num{3.03e9} & \num{0.00e0} & \num{1.31e7} & \num{9.75e7} & \num{5.70e8} & \num{4.61e10} \\
Gross Profit Margin & \num{1.30e4} & \num{2.75e1} & \num{2.03e1} & \num{-3.25e2} & \num{1.38e1} & \num{2.34e1} & \num{3.77e1} & \num{1.15e2} \\
Intangible Assets & \num{1.37e4} & \num{2.57e9} & \num{1.09e10} & \num{-1.78e8} & \num{1.10e8} & \num{3.78e8} & \num{1.32e9} & \num{2.69e11} \\
Interest Expense & \num{3.99e2} & \num{6.51e8} & \num{2.37e9} & \num{0.00e0} & \num{4.69e6} & \num{3.99e7} & \num{1.79e8} & \num{1.92e10} \\
MarketValue & \num{2.98e6} & \num{6.04e10} & \num{1.68e11} & \num{1.21e8} & \num{1.04e10} & \num{2.23e10} & \num{4.72e10} & \num{8.05e12} \\
Net Income & \num{1.40e4} & \num{3.69e9} & \num{1.87e10} & \num{-6.87e10} & \num{1.14e8} & \num{5.39e8} & \num{1.70e9} & \num{3.67e11} \\
Operating Costs & \num{1.40e4} & \num{3.51e10} & \num{1.43e11} & \num{-5.74e7} & \num{2.19e9} & \num{6.51e9} & \num{2.07e10} & \num{3.23e12} \\
Operating Margin & \num{1.40e4} & \num{4.00e-3} & \num{1.19e1} & \num{-1.11e3} & \num{2.75e-2} & \num{8.94e-2} & \num{2.07e-1} & \num{1.81e2} \\
Operating Profit & \num{1.40e4} & \num{4.56e9} & \num{2.30e10} & \num{-7.16e10} & \num{1.22e8} & \num{6.26e8} & \num{1.99e9} & \num{4.24e11} \\
Operating Revenue & \num{1.40e4} & \num{3.93e10} & \num{1.54e11} & \num{-3.87e8} & \num{2.49e9} & \num{7.23e9} & \num{2.31e10} & \num{3.32e12} \\
Paid-in Capital & \num{1.40e4} & \num{5.29e9} & \num{2.44e10} & \num{4.20e7} & \num{6.99e8} & \num{1.42e9} & \num{3.14e9} & \num{4.62e11} \\
R\&D Expenses & \num{5.03e3} & \num{1.08e9} & \num{3.09e9} & \num{0.00e0} & \num{6.72e7} & \num{2.83e8} & \num{8.12e8} & \num{5.32e10} \\
ROE & \num{1.39e4} & \num{6.25e0} & \num{1.06e2} & \num{-7.78e3} & \num{3.61e0} & \num{8.77e0} & \num{1.50e1} & \num{1.35e3} \\
ROIC & \num{1.40e4} & \num{2.95e-2} & \num{1.24e0} & \num{-1.14e2} & \num{6.51e-3} & \num{3.94e-2} & \num{9.03e-2} & \num{5.13e1} \\
Revenue & \num{1.40e4} & \num{3.94e10} & \num{1.54e11} & \num{-3.87e8} & \num{2.53e9} & \num{7.36e9} & \num{2.33e10} & \num{3.32e12} \\
Shareholders Equity & \num{1.40e4} & \num{3.61e10} & \num{1.62e11} & \num{-2.91e10} & \num{2.87e9} & \num{7.44e9} & \num{1.91e10} & \num{3.99e12} \\
Total Assets & \num{1.40e4} & \num{2.51e11} & \num{1.84e12} & \num{0.00e0} & \num{5.66e9} & \num{1.56e10} & \num{4.61e10} & \num{4.88e13} \\
Total Liabilities & \num{1.40e4} & \num{2.15e11} & \num{1.69e12} & \num{0.00e0} & \num{2.13e9} & \num{7.27e9} & \num{2.74e10} & \num{4.48e13} \\
\bottomrule
\end{tabular}
    }
\end{table}

\section{Within the Macro Agent: Merrill Lynch Clock vs Industry Momentum}\index{Macro agent inside weighting}
During the training period, the Merrill Lynch Clock:Industry Momentum configuration with a 25:75 weighting was selected for the Macro agent, as it optimally integrates macroeconomic and industry momentum scores. This configuration was chosen based on its superior risk-adjusted performance, demonstrated by the highest combined Sharpe and Sortino ratios (2.514) among the evaluated configurations (see Table~\ref{tab:Merrill Lynch Clock}). Consequently, this setting was adopted for out-of-sample testing to assess its generalizability.

\begin{table}[ht]
\centering
\caption{Within the Macro Agent: Merrill Lynch Clock vs Industry Momentum}
\label{tab:Merrill Lynch Clock}
\resizebox{\textwidth}{!}{%
\begin{tabular}{lrrrrrrrr}
\toprule
\textbf{Model} & \textbf{CR (\%)} & \textbf{AR (\%)} & \textbf{STD (\%)} & \textbf{DD (\%)} & \textbf{Sharpe} & \textbf{Sortino} & \textbf{MDD (\%)} & \textbf{Calmar} \\
\midrule
Merrill Lynch Clock:Industry Momentum = 25:75 & 89.30 & 17.86 & 22.40 & 10.40 & \textbf{0.797} & \textbf{1.717} & -41.00 & 0.436 \\
Merrill Lynch Clock:Industry Momentum = 50:50 & 88.79 & 17.76 & 22.57 & 11.11 & 0.787 & 1.599 & -41.84 & 0.424 \\
Merrill Lynch Clock:Industry Momentum = 75:25 & 63.22 & 12.65 & 15.65 & 9.25 & 0.808 & 1.367 & -16.48 & 0.767 \\
\bottomrule
\end{tabular}%
}
\end{table}

\section{Notation}

\begin{table}[ht!]
\centering
\caption{Key Notation for Problem Formulation} \label{tab:notation}
\begin{tabular}{@{}ll@{}}
\toprule
Symbol & Description \\
\midrule
\(N, T\) & Number of assets, number of trading days \\
\(d_m, d_f, d_s, d_b\) & Dimensions of macro-economic, firm, industry, benchmark features \\
\(\mathbf{M}, \mathbf{F}, \mathbf{S}, \mathbf{B}\) & Macro-, Industry-, Firm-, and benchmark-level data tensors/matrices \\
\(\mathbf{R}\) & Matrix of realized asset returns \\
\(\mathbf{w}_t\) & Portfolio weights at time \(t\) \\
\(r_t, x_t\) & Portfolio return and excess return at time \(t\) \\
\(r_f\) & Risk-free rate \\
\(f_\theta\) & Parameterized allocation rule with parameters \(\theta\) \\
\(\mathrm{SR}(\theta)\) & Sharpe Ratio of the strategy induced by \(\theta\) \\
\bottomrule
\end{tabular}

\end{table}

\end{document}